\documentclass{article}
\usepackage{spconf,amsmath,graphicx}
\usepackage{multirow}
\usepackage{graphicx}
\usepackage{amsmath}
\usepackage[psamsfonts]{amssymb}
\usepackage{amsxtra}
\usepackage[T1]{fontenc}
\usepackage{threeparttable}
\usepackage{epsfig,amssymb,amsmath}
\usepackage{multirow}
\usepackage{subfigure}
\usepackage{mathrsfs}
\usepackage{float}
\usepackage{hyperref}
\usepackage{url}
\usepackage[ruled,vlined]{algorithm2e}
\usepackage{cite}
\usepackage{xcolor}

\title{VAW-GAN for Disentanglement and Recomposition of Emotional Elements in Speech}
\name{Kun Zhou\thanks{\textbf{Speech Samples:} \url{https://demo9646.github.io/slt2020_demo/}} $^1$, Berrak Sisman $^{2}$, Haizhou Li $^1$}
\address{
$^1$ Department of Electrical and Computer Engineering, National University of Singapore\\
$^2$ Information System Technology and Design, Singapore University of Technology and Design}

\begin{document}
%
\topmargin=0mm
\maketitle
\begin{abstract}
Emotional voice conversion (EVC) aims to convert the emotion of speech from one state to another while preserving the linguistic content and speaker identity. In this paper, we study the disentanglement and recomposition of emotional elements in speech through variational autoencoding Wasserstein generative adversarial network (VAW-GAN). We propose a speaker-dependent EVC framework based on VAW-GAN, that includes two VAW-GAN pipelines, one for spectrum conversion, and another for prosody conversion. We train a spectral encoder that disentangles emotion and prosody (F0) information from spectral features; we also train a prosodic encoder that disentangles emotion modulation of prosody (affective prosody) from linguistic prosody. At run-time, the decoder of spectral VAW-GAN is conditioned on the output of prosodic VAW-GAN. The vocoder takes the converted spectral and prosodic features to generate the target emotional speech.  Experiments validate the effectiveness of our proposed method in both objective and subjective evaluations.
\end{abstract}
\section{Introduction}
\begin{figure}[t]
    \centering
    \includegraphics[width=70mm]{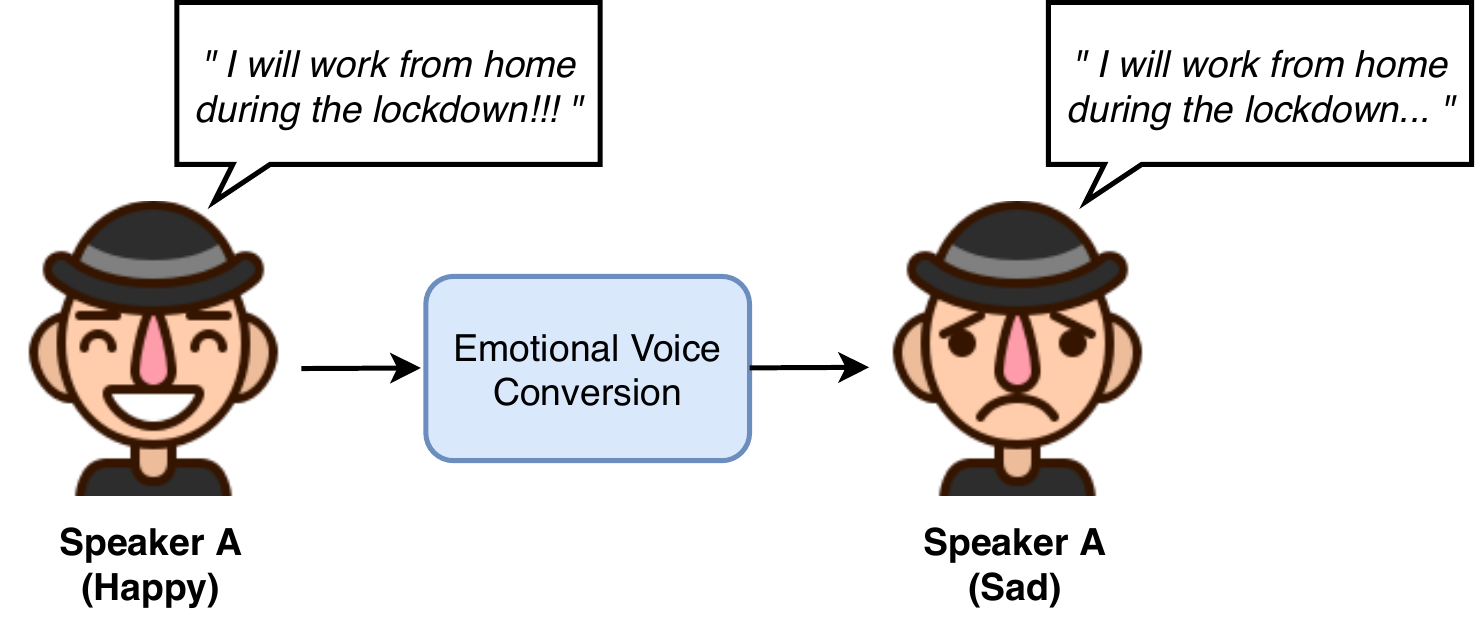}
    \vspace{-3mm}
    \caption{An emotional voice conversion system is trained with  speech data in different emotional patterns from the same speaker. At run-time, the system takes the speech of one emotion as the input, and converts to that of another \cite{ Zhou2020,shankar2019automated,zhou2020converting}.}
    \label{fig:em-conversion}
    \vspace{-6mm}
\end{figure}

Speech conveys information not only with lexical words but also through its prosody. Speech prosody can affect the syntactic and semantic interpretation of an utterance \cite{hirschberg2004pragmatics}, that is called linguistic prosody. It also displays one's emotional state, that is referred to as emotional prosody~\cite{arnold1960emotion}. Emotional voice conversion is a voice conversion (VC) technique to convert the emotional prosody of speech from one to another while preserving the linguistic content and speaker identity, as shown in Figure \ref{fig:em-conversion}. EVC is an enabling technology for many applications such as text-to-speech \cite{liu2020teacher, liu2020wavetts,liu2020expressive}, personalized speech synthesis \cite{tits2020neural,tits2019visualization} and conversational robots \cite{tits2020ice}.

In general, voice conversion aims to change the speaker identity of speech while preserving the linguistic content \cite{sisman2020overview}. The early studies of VC include Gaussian mixture model (GMM) \cite{toda2007voice}, partial least square regression~ \cite{helander2010voice} and sparse representation \cite{ccicsman2017sparse, sisman2018phonetically, sisman2018voice,sisman2019group}. Recent deep learning approaches, such as deep neural network (DNN) \cite{chen2014voice, lorenzo2018investigating}, recurrent neural network (RNN) \cite{nakashika2014high} and generative adversarial network (GAN) \cite{sismanstudy} have advanced the state-of-the-art. To eliminate the need for parallel training data, variational autoencoding network (VAE) \cite{hsu2016voice}, cycle-consistent generative adversarial network (CycleGAN) \cite{kaneko2017parallel} and star generative adversarial network (StarGAN) \cite{kameoka2018stargan} have been proposed.



As  speaker identity is thought to be characterised by  physical attributes of  speaker and strongly affected by  spectrum \cite{xu2011speech}, spectral mapping has been the main focus of conventional voice conversion \cite{sisman2020overview}. However, we note that emotion is inherently supra-segmental and hierarchical in nature, which is highly complex with multiple attributes entangled with the spectrum and prosody \cite{xu2011speech, latorre2008multilevel,schuller2020review}. Emotional prosody is perceived in different time scales at phoneme, word, and sentence level ~\cite{csicsman2017transformation,sanchez2014hierarchical}. Therefore, it cannot be analyzed simply at the frame level, nor can it be simply converted from the spectrum alone. Prosody conversion plays an important role in emotional voice conversion, which will also be studied in this paper. 

Some earlier studies on emotional voice conversion use the logarithm Gaussian (LG)-based linear transformation method \cite{gao2019nonparallel,tao2006prosody, wu2009hierarchical} to convert F0. Such simple representation of F0 is insufficient to characterize the prosody in different time scales~\cite{latorre2008multilevel,sisman2019group}. Continuous wavelet transform (CWT)  describes utterance level prosody with a multi-resolution representation~\cite{csicsman2017transformation,sisman2018wavelet}, which allows us to handle prosody conversion at different temporal scales~\cite{Zhou2020, sisman2019group}. We will continue to explore the use of CWT coefficients of F0 as prosodic features in this paper.

Statistical modelling techniques are simple and effective in prosody conversion. One idea is to use a classification or regression tree \cite{tao2006prosody, wu2009hierarchical} to decompose the pitch contour of the source speech into a hierarchical structure, that is followed by GMM and regression-based clustering methods. Another strategy was to create a source and target dictionary for both spectral and prosodic features, and estimate a sparse mapping using exemplar-based techniques with NMF \cite{aihara2014exemplar}. Moreover, there was also a study to combine hidden Markov model (HMM),  GMM,  and  F0  segment selection \cite{inanoglu2009data} for spectrum and prosody conversion. Recent deep learning methods, such as deep belief network (DBN) \cite{luo2016emotional}, deep bi-directional long-short-term memory (DBLSTM) \cite{ming2016deep}, highway neural network \cite{shankar2019automated,shankar2019multi}, sequence-to-sequence \cite{robinson2019sequence} and rule-based model \cite{xue2018voice} have achieved remarkable performance on emotion conversion. We note that the prior studies of emotion conversion do not provide an in-depth investigation of the disentanglement of emotional elements in speech, which will be the focus of this paper. 





We note that CycleGAN is an effective approach for prosody mapping and has already been used in emotional voice conversion \cite{ Zhou2020}. CycleGAN performs one-to-one emotion conversion, and do not analyze the emotional factors of speech through disentanglement.  An encoder-decoder structure, such as VAW-GAN \cite{hsu2017voice}, would be more suitable as its encoder learns to disentangle emotion information from others, and generate a latent code from an emotion-independent distribution. By conditioning on controllable attributes, such as emotion labels and F0 values, the decoder can recompose speech of new emotion types, that facilitates many-to-many emotional voice conversion.

The main contributions of this paper include: 1) we propose an emotional voice conversion framework with VAW-GAN that is trained on non-parallel data; 2) we study the use of CWT decomposition to characterize F0 for VAW-GAN prosody mapping; and 3) we propose to use an encoder to disentangle emotion information from speech content, and a decoder to recompose target emotional speech.

This paper is organised as follows: In Section 2, we motivate our study and introduce the related work. In Section 3, we propose an emotional voice conversion framework that disentangles the emotional factors of speech through VAW-GAN. Section 4 reports the objective and subjective evaluation. Section 5 concludes the study.  

\section{Motivation and Related Work}
\subsection{Spectrum Conversion with VAW-GAN}


Variational autoencoder (VAE) has been adopted in voice conversion with non-parallel training data~\cite{huang2019investigation,hsu2016voice,Qian_2020,huang2019refined}. It consists of an encoder and a decoder, where an encoder $E_{\phi}$ with parameter set $\phi$ is similar to a phone recognizer to encode the input $x$ to a latent code, and a decoder $G$ with parameter set $\theta$ operates as a synthesizer to reconstruct the input $x$. 


Let $\textbf{x}_s$ be the spectral frames from  source speaker and  $\textbf{x}_t$ from target speaker.
During training, the encoder $E_{\phi}$ encodes an input spectral frame $x$ into a latent code: $z = E_{\phi}(x)$, and the decoder $G_{\theta}$ reconstructs the input $x$ with the corresponding speaker code $y$. The reconstructed input $\hat{x}$ is given as:  
\begin{equation}
    \hat{x} =G_{\theta}(z,y) = G_{\theta}(E_{\phi}(x),y)
\label{eq:1}
\end{equation}
The model parameters $\theta$ can be obtained by maximizing the variational lower bound:
\begin{multline}
    J_{vae}(x|y)\\ = - D_{KL}(q_{\phi}(z|x)\Vert p_{\theta}(z)) + \mathbb{E}_{q_{\phi}(z|x)}[ \log p_{\theta}(x|z,y) ]
\end{multline}
where $x \in \textbf{x}_s \cup \textbf{x}_t$, $D_{KL}$ is the Kullback-Leibler divergence, $q_{\phi}(z|x)$ is the approximate posterior, $p_{\theta}(x|z,y)$ is the data likelihood, and $p_{\theta}(z)$ is the prior distribution of the latent space.
\begin{figure}[t]
    \centering
    \includegraphics[width=7cm]{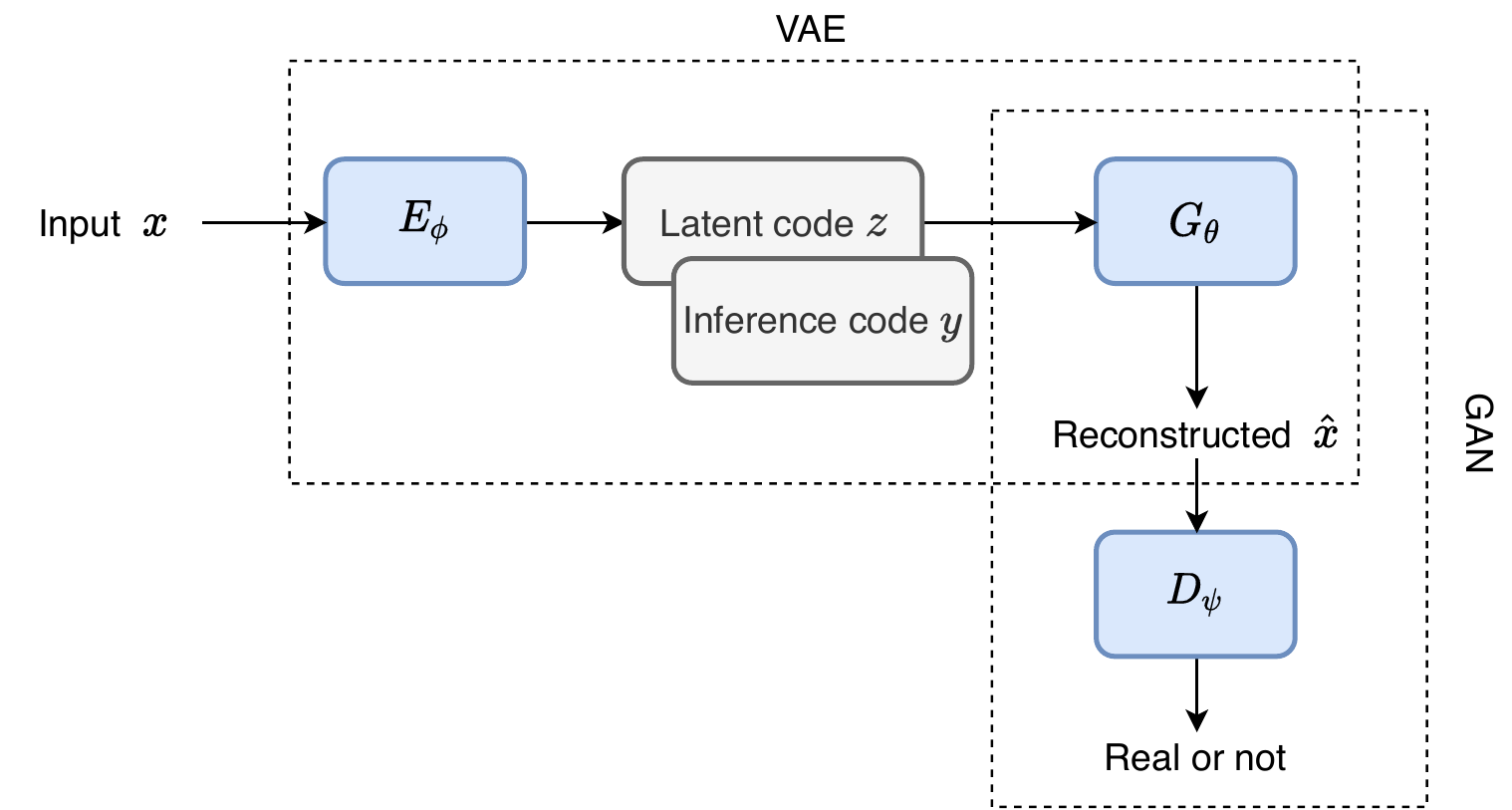}
    \vspace{-3mm}
    \caption{Illustration of the VAW-GAN framework. A VAW-GAN is incorporated with a VAE and a GAN model by assigning VAE's decoder as GAN's generator.}
    \label{fig:vawgan}
    \vspace{-3mm}
\end{figure}

\begin{figure*}[t]
    \centering
    \includegraphics[width=14cm]{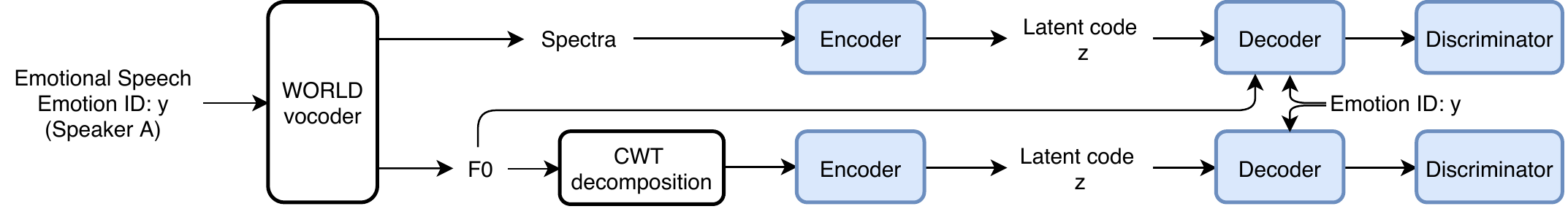}
    \vspace{-2mm}
    \caption{The training phase of the proposed VAW-GAN-based EVC framework with WORLD vocoder. Blue boxes are involved in the training, while white boxes are not. }
    \vspace{-4mm}
    \label{fig:training}
\end{figure*}

VAE is effective in disentanglement of mixed attributes without the need of parallel training data~\cite{hsu2017voice,mohammadi2018investigation}. Recently, generative adversarial network (GAN)-based model has been successfully used in many machine learning applications such as image synthesis \cite{ak2020incorporating, emir2019semantically, ak2020semantically}, image manipulation \cite{icipgan, ak2019deep, ak2019attribute}, speech synthesis \cite{kaneko2017generative, zhao2018wasserstein} and voice conversion \cite{singan-2019,berrak_ganslt, guo2019new}. A GAN consists of two main components: a generator $G_{\theta}$ that generates the realistic spectrum and a discriminator $D_{\psi}$ that decides whether the input is real or generated. It is noticed that GAN models can produce sharper spectra by optimizing a loss function between two distributions through a min-max game. To benefit from VAE and GAN, VAW-GAN \cite{hsu2017voice} is proposed for voice conversion, which combines the objectives of VAE and W-GAN \cite{arjovsky2017wasserstein} by assigning VAE's decoder as GAN's generator, as illustrated in Figure \ref{fig:vawgan}. 

Given the target speaker representation $y_t$, the final objective is given as:
\begin{multline}
    J_{vawgan}(\phi,\theta, \psi;x,y,z)\\ = - D_{KL}(q_{\phi}(z|x)\Vert p(z))
    + \mathbb{E}_{q_{\phi}(z|x)}[ \log p_{\theta}(x|z,y)\\
    + \alpha \mathbb{E}_{x\sim p_t^*}[D_{\phi}(x)]
    + \alpha \mathbb{E}_{z\sim q_{\psi}(z|x)}[D_{\psi}(G_{\theta}(z,y_t))]
\end{multline}
where $\alpha$ is a trade-off parameter. This objective is shared across all three main components in VAW-GAN: the encoder, the decoder/generator and the discriminator. During the training, the decoder/generator tries to minimize the loss whereas the discriminator maximizes it. 

\subsection{Prosody Characterization with Wavelet Transform}
Emotional prosody usually refers to pitch, intonation, intensity, and speaking rate at segmental and supra-segmental level. F0 is used to quantitatively describe pitch and intonation. However, modelling F0 is challenging due to its discontinuity and hierarchy in nature \cite{wang2019vector,zhao2020improved}. The prosody of an utterance is a modulation between linguistic prosody and emotional prosody. As far as emotion conversion is concerned, linguistic prosody of an utterance is emotion independent. Converting the emotion, we would like to carry over the linguistic prosody from source to target, but change source emotional prosody to target one. 

We hypothesize that linguistic prosody and emotional prosody are manifested in different time scales. A multi-scale F0 decomposition allows us to characterize and manipulate F0 in a more effective way. CWT is a multi-scale modelling technique, that decomposes F0 into different temporal scales. Such multi-scale CWT coefficients are used successfully in speech synthesis \cite{kruschke2003estimation,mishra2006decomposition} and voice conversion \cite{sisman2019group, sisman2018wavelet}.
Following the same idea, we decompose F0 into 513 scales, varying from the micro-prosody level to the whole utterance level. In the 513-scale representation, the fine scale coefficients capture the short-term variations, while the coarse scale coefficients capture the long-term variations. We adopt Mexican hat as the mother wavelet and choose translating factor $\tau_0$ as 5 ms. 
We use the CWT coefficients of F0 as the prosodic features as will be further discussed in Section III. 



\begin{figure*}[t]
    \centering
    \includegraphics[width=16cm]{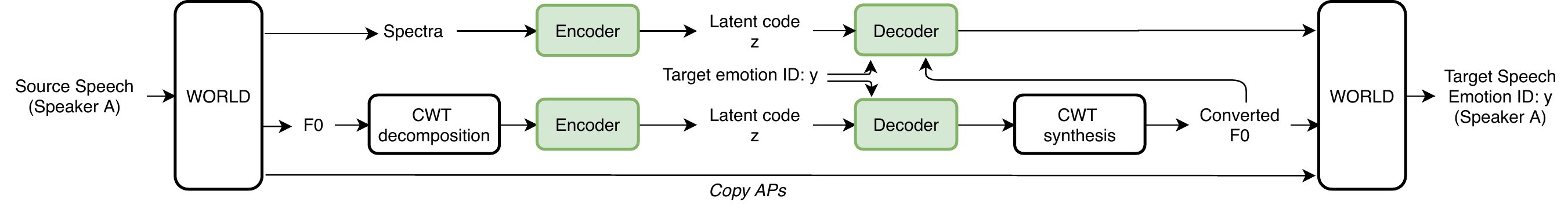}
    \vspace{-4mm}
    \caption{The run-time conversion phase of the proposed VAW-GAN-based EVC framework. Green boxes represent the networks that are already trained.}
    \label{fig:conversion}
    \vspace{-4mm}
\end{figure*}

\begin{figure}[ht]
    \centering
    \includegraphics[width=7cm]{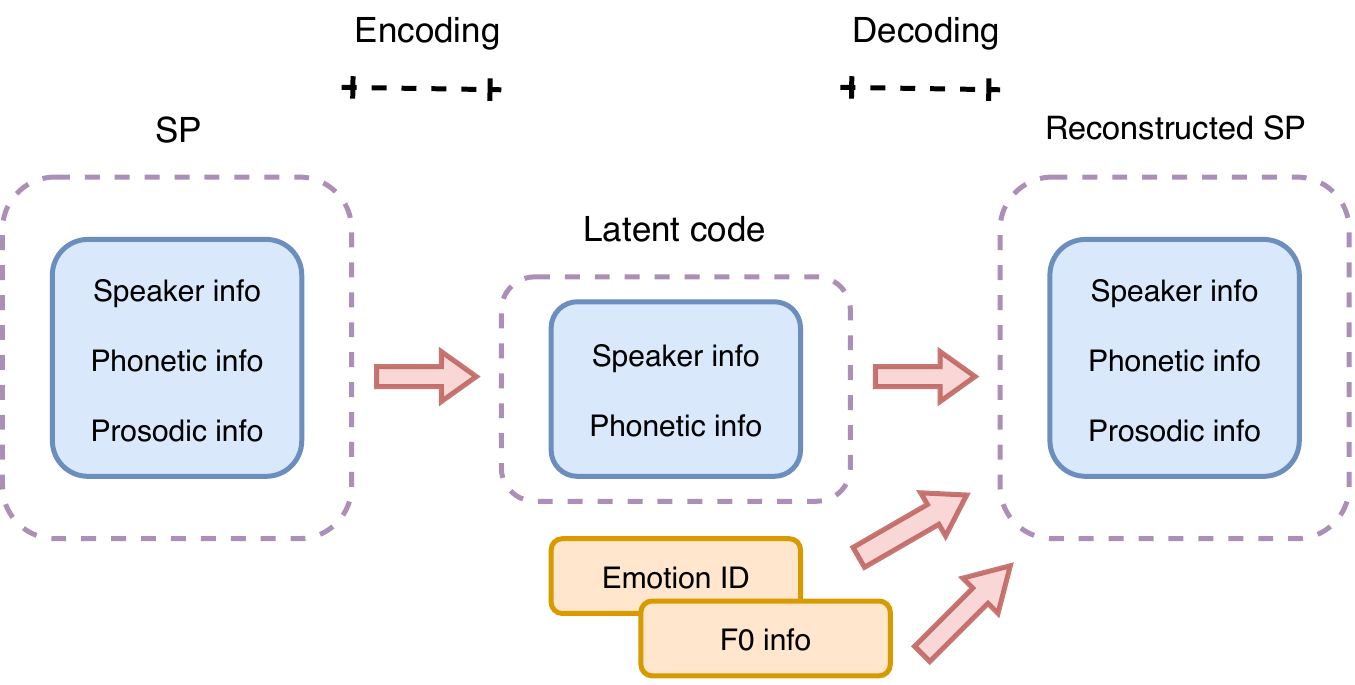}
    \vspace{-3mm}
    \caption{Disentanglement of the emotion-independent latent code with conditional variables. By providing the emotion ID and F0 explicitly, the encoder learns to discard emotion-related information during training. }
    \label{fig:condition}
    \vspace{-6mm}
\end{figure}

\section{Disentanglement of Emotional Elements for EVC}

In this section, we propose to disentangle emotional elements of speech through VAW-GAN. To achieve this, we propose a speaker-dependent EVC framework, that consists of two parallel pipelines for spectrum and prosody conversion, as shown in Figure \ref{fig:training}.  Each VAW-GAN pipeline is trained so that the encoder disentangles emotion elements, and the decoder is conditioned on both emotion type and F0 to recompose the emotional speech with target emotion. The proposed framework is referred to as \textit{VAW-GAN (SP+CWT+C)}, where `SP' and `CWT' denote the two parallel pipelines, and `C' denotes that the decoder is conditioned on CWT coefficients.

We first extract spectral features (SP) and F0 features from the raw audio using the WORLD vocoder \cite{morise2016world}. 
We perform CWT decomposition on F0 to describe the prosody from the micro-prosody level to the whole utterance level. The CWT decomposition of F0 allows the encoder to learn the emotional patterns in different time scales. It is noticed that F0 obtained from the vocoder is discontinuous, due to the voiced/unvoiced parts in the speech. Therefore, the following pre-processing procedures are necessary: 1) linear interpolation over unvoiced regions, 2) transformation of  F0 from linear to a logarithmic scale, and  3)  normalization of the resulting F0 to zero mean and unit variance.
\subsection{Training Phase}
The training phase of the proposed framework is shown in Figure \ref{fig:training}. 
The spectral encoder learns to disentangle emotion and F0 from speech content. The resulting latent code is emotion-independent. The prosodic encoder learns to disentangle affective prosody from linguistic prosody. The resulting latent code is emotion-independent. 
The earlier study \cite{Zhou2020} has shown that  separate training of spectrum and prosody achieves better performance than  joint training. Following this idea, we propose to train two VAW-GAN-based networks separately: 1) a VAW-GAN model conditioned on F0 for spectrum conversion, denoted as \textit{VAW-GAN for Spectrum}, and 2) a VAW-GAN model with CWT analysis for prosody conversion, which is denoted as \textit{VAW-GAN for Prosody}.

Auto-encoder is often used to learn disentangled representations. We expect that \textit{VAW-GAN for prosody}  learns the prosodic pattern in different time-scales, and \textit{VAW-GAN for spectrum} disentangles the phonetic and speaker information from  emotion-related prosodic information, and represents them in the latent code.

Both networks contain three main components, that are: 1) encoder, 2) decoder/generator, and 3) discriminator. During the training, the encoder is exposed to input frames from the same speaker but with different emotions. The encoder learns the emotion-independent patterns and transforms the input frames into a latent code $z$. We assume that the latent code $z$ only contains the information of speaker identity and phonetic content. An one-hot vector emotion ID is used to provide the emotion information to the decoder/generator. Since the spectral features obtained from the vocoder are highly dependent on F0 and contain prosodic information, we propose to add F0 as an additional input to the decoder/generator during the training of \textit{VAW-GAN for Spectrum}. In this way, we assume that \textit{VAW-GAN for spectrum} could eliminate the prosodic information in the spectral features and generate an emotion-independent latent code $z$ during the training.

We then train a generative model for spectrum and prosody through an adversarial training: the discriminator tries to maximize the loss between the real and reconstructed features, while the generator is used to minimize it. The generative model tries to find an optimal solution through this min-max game, which allows us to generate high-quality speech samples.
\vspace{-4mm}
\subsection{Run-time Conversion}
The run-time conversion phase is shown in Figure \ref{fig:conversion}. For prosody conversion, we perform CWT on F0 to decompose it into different time scales. The CWT-based F0 features are then converted by the trained \textit{VAW-GAN for prosody} with the designated emotion ID to generate the converted CWT-based F0 features. As for spectrum conversion, we propose to condition the decoder/generator on the converted CWT-based F0 features together with the designated emotion ID. Then the converted spectral features are obtained through the trained \textit{VAW-GAN for spectrum}. Finally, the converted speech is synthesized by WORLD vocoder with the converted spectral and prosody features. It is noted that aperiodicities (APs) are directly copied from the source speech.

\subsection{Conditioning on F0 for Decoding }
It is well known that both WORLD \cite{morise2016world} and STRAIGHT \cite{kawahara1999restructuring} vocoder extract the spectral features from the F0 during feature extraction, which makes the spectral features contain prosodic information. Therefore, in some VAE-based voice conversion (VAE-VC) frameworks, the latent code extracted from the source spectral features still contains the information of the source F0, and the conversion performance may suffer from this unwanted entanglement. In speech synthesis, multi-tier approaches to speech representation, where parameters modeled at one tier are used to condition the modelling of the following tier, are proposed for waveform generation \cite{wang2017rnn,wang2018comparison,watts2019speech}. Inspired by that, some VAE-VC frameworks \cite{huang2019investigation, Qian_2020} propose to use F0 as an additional condition variable override the F0 information in the latent code.

During the training, given the $F_0$ features extracted from the speech, the reconstructed input $\overline{x}$ from the equation (\ref{eq:1}) can be modified as:
\begin{equation}
    \overline{x} = G_{\theta}(z,y,F_0) = G_{\theta}(E_{\phi}(x),y, F_0) 
    \label{eq:4}
\end{equation}

In the conversion phase, given the converted F0 features $\hat F_0$, the converted features $\hat{x}$ from the equation (\ref{eq:4}) can be modified as:
\begin{equation}
    \hat x = G_{\theta}(z, y_t, \hat F_0) = G_{\theta}(E_{\phi}(x), y_t, \hat F_0)
\end{equation}
where $y_t$ is the designated target emotion code.

As illustrated in Figure \ref{fig:condition}, the spectral features are highly dependent on F0 and contain prosodic information, it is insufficient to train an emotion-independent encoder only with one-hot vector emotion ID. Therefore, in both training and conversion phase, we propose to condition F0 features to the decoder. In this way, the encoder could learn to discard emotion-related information during training and generate an emotion-independent representation, which allows a more effective control of the output prosody.  


\section{Experiments}

We conduct experiments on EmoV-DB \cite{adigwe2018emotional}, which is an English speech emotion corpus. During training, we choose one male speaker, denoted as \textit{Sam}; and one female speaker, denoted as \textit{Bea} from EmoV-DB. The proposed framework is trained with 300 non-parallel utterances from each speaker and no time-alignment is required. We choose three emotions that are 1) neutral, 2) angry and 3) sleepy. In all experiments,we conduct emotion conversion from 1) neutral to angry (N2A), and 2) neutral to sleepy (N2S). At run-time, we use 2 minutes of speech data to perform objective and subjective evaluation. 
\subsection{Baseline Frameworks}
To evaluate the effectiveness of our proposed framework, denoted as \textit{VAW-GAN (SP+CWT+C)}, we have implemented two baseline systems: 1) \textit{VAW-GAN (SP+CWT)} that is VAW-GAN based EVC framework, where both spectral and CWT-based F0 features are converted without F0 conditioning on the spectral decoder; and 2) \textit{VAW-GAN (SP+F0+C)} that is VAW-GAN based EVC framework, where both spectral features are converted with  LG-based F0 conditioning on the spectral decoder, and F0 is converted with LG-based linear transformation. 
\begin{table}[t]
\centering
\caption{MCD [dB] and LSD [dB] results for two comparative systems: (1) \textit{VAW-GAN (SP+CWT+C)}; (2)\textit{VAW-GAN (SP+CWT)};  Notes: M: male speaker; F: female speaker.}
\scalebox{0.9}{
\begin{tabular}{c||c|c||c|c}
\hline
\multirow{2}{*}{VAW-GAN} & \multicolumn{2}{c||}{N2A} & \multicolumn{2}{c}{N2S} \\ \cline{2-5} 
                           & MCD          & LSD      & MCD       & LSD       \\ \hline
(SP+CWT) (M)                  & 3.626       & 5.070      & 3.650       & 5.231      \\ 
\textbf{(SP+CWT+C) (M)}               & 3.238       & 4.558      & 3.587       & 5.094      \\ 
(SP+CWT) (F)  & 4.127 & 5.831 & 4.350 &6.209  \\
\textbf{(SP+CWT+C) (F)} & 4.085 & 5.681 & 4.278 & 6.106 \\
\hline
\end{tabular}}
\label{table:mcd}
\vspace{-4mm}
\end{table}
\subsection{Experimental Setup}
In our proposed framework, we train two VAW-GAN pipelines for spectrum and prosody, as shown in Figure \ref{fig:training}. For both pipelines, the encoder is a 5-layer 1D convolutional neural network (CNN) with the kernel size of 7 and a stride of 3 followed by a fully connected layer. Its output channel is $\{16,32,64,128,256\}$, and the latent code is 128-dimensional which is assumed to have a standard normal distribution. In prosody pipeline, the emotion ID is a 10-dimensional one-hot vector and concatenated with the latent code $z$ to form a 138-dimensional vector, then merged by a fully connected layer. In spectrum pipeline, one-dimensional F0 is concatenated together with the latent code $z$ and the 10-dimensional emotion ID and merged as the input to the decoder. The decoder is 4-layer 1D CNN with kernel sizes of $\{9,7,7,1025\}$ and  strides  of $\{3,3,3,1\}$. Its output channel  is $\{32,16,8,1\}$. The discriminator is a 3-layer 1D CNN with kernel sizes of $\{7,7,115\}$ and strides of $\{3,3,3\}$ followed by a fully connected layer. We train the networks by using RMSProp with the learning rate of 1e-5, and the batch size is set as 256 for 45 epochs.

In all experiments, speech data are sampled at 16 kHz with 16-bit per sample. The FFT length is 1024, and 513-dimensional spectral features (SP), fundamental frequency (F0) and AP are extracted every 5 ms using WORLD vocoder. The frame length is 25 ms with a frame shift of 5 ms. In the pre-processing step, we normalize every input frame of SP to unit sum and the normalizing factor (energy) is taken out as an independent feature. We further re-scale SP to logarithm and re-scale log energy-normalized SP to the range of [-1, 1] per dimension. It is noted that we carry over the energy and AP components from source to target without modification.

\subsection{Objective Evaluation}

\subsubsection{Spectrum Conversion}

We employ Mel-cepstral distortion (MCD) and log spectral distortion (LSD) to assess the performance of spectrum conversion. MCD measures the distortion between the converted and target Mel-cepstra.
In this paper, we extract 24-dimensional MCEPs at each frame. It is noted that a lower MCD indicates a better performance.
Moreover, LSD is used to measure the distortion between the converted and target spectral envelope using Euclidean distance between the feature vectors. 
In this paper, we coded the spectral features into 24-dimension with WORLD vocoder. A
smaller LSD value represents lower spectral distortion, thus  better performance.

\begin{table}[t]
\centering
\caption{RMSE [Hz] and PCC results for: (1) \textit{VAW-GAN (SP+CWT+C)}; (2)\textit{VAW-GAN (SP+F0+C)}.}
\scalebox{0.9}{
\begin{tabular}{c||c|c||c|c}
\hline
\multirow{2}{*}{Framework} & \multicolumn{2}{c||}{N2A} & \multicolumn{2}{c}{N2S} \\ \cline{2-5} 
                           & RMSE         & PCC       & RMSE        & PCC     \\ \hline
(SP+F0+C) (M)                  & 42.331      & 0.887      & 46.587       & 0.836     \\ 
\textbf{(SP+CWT+C) (M)}             & 40.440       & 0.903      & 43.905       & 0.844      \\ 
(SP+F0+C) (F)       & 67.124 & 0.868 & 54.824 & 0.859        \\
\textbf{(SP+CWT+C) (F)} & 61.712 & 0.893 & 52.348 &0.867 \\
\hline
\end{tabular}}
\label{table:pcc}
\vspace{-4mm}
\end{table}
Table 1 reports the MCD and LSD values for spectrum conversion for 1) neutral-to-angry and 2) neutral-to-sleepy. We observe that the proposed  \textit{VAW-GAN (SP+CWT+C)} framework consistently outperforms \textit{VAW-GAN (SP+CWT)}, where the decoder does not  condition on F0 information. 
We note that for both male and female speakers, we achieve remarkable and consistent performance.  The results suggest that F0 conditioning mechanism has an positive effect on spectrum conversion by reducing the spectrum distortion. 

\subsubsection{Prosody Conversion}
We use root mean square error (RMSE) \cite{luo2016emotional} and Pearson correlation coefficient (PCC) \cite{sisman2019group} to report the performance of prosody conversion. 
A lower RMSE indicates a better F0 conversion performance.
The PCC measures the linear dependency between the converted and target interpolated F0 sequences.
We note that a higher PCC value represents a stronger linear dependency and better F0 conversion performance.

Table 2 reports the RMSE and PCC results to assess the prosody conversion performance. In this experiment, we conducted two emotion conversion settings: N2A and N2S, similar to that of the spectrum evaluation. We compare the proposed framework \textit{VAW-GAN (SP+CWT+C)} with the baseline framework~\textit{VAW-GAN (SP+F0+C)}, and observe  that \textit{VAW-GAN (SP+CWT+C)} outperforms the baseline. The results suggest that VAW-GAN with CWT decomposition of F0 is  more effective than LG-based F0 conversion.

\subsection{Subjective Evaluation}
We further conduct three listening experiments to assess the proposed framework in terms of speech quality and emotion similarity. 11 subjects participated in all the experiments and each of them listened to 40 utterances in total. We perform XAB test to assess the speech quality and emotion similarity, which has been widely used in speech synthesis, such as voice conversion \cite{berrak_ganslt}, singing voice conversion \cite{singan-2019}.
\begin{figure}[t]
    \centering
    \includegraphics[width=6cm]{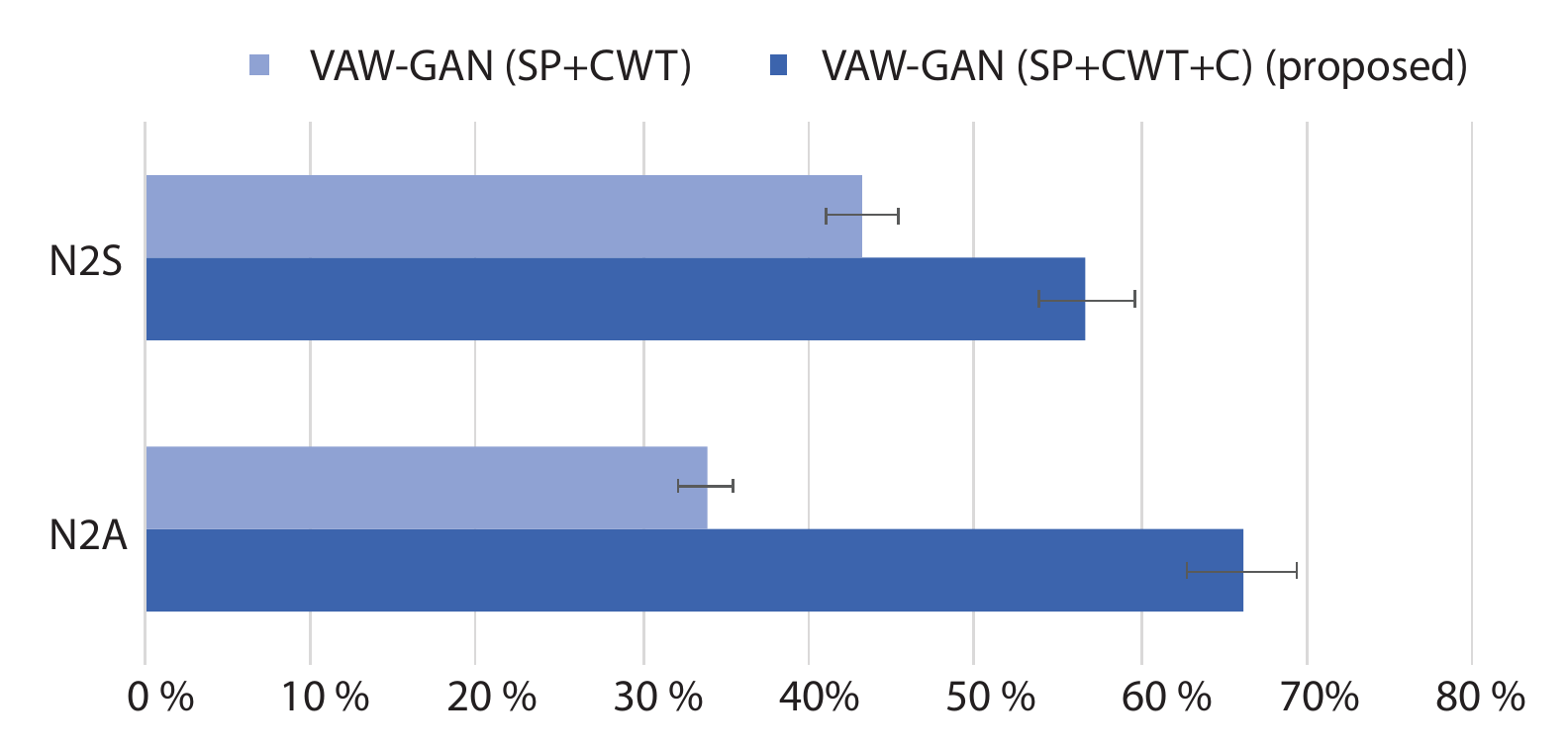}
    \vspace{-4mm}
    \caption{XAB speech quality preference results with 95 $\%$ confidence interval to assess the effect of F0 conditioning.}
    \vspace{-3mm}
    \label{fig:ab1}
\end{figure}

We first conduct XAB test between our proposed framework \textit{VAW-GAN (SP+CWT+C)} and the baseline framework \textit{VAW-GAN (SP+CWT)} to assess the performance of speech quality, where subjects are asked to choose the one with better speech quality. As shown in Figure \ref{fig:ab1}, we observe that our proposed framework consistently outperform the baseline framework for both N2S and N2A. We note that both frameworks convert F0 with CWT, whereas the proposed framework conditions the decoder with F0 features and the baseline framework does not have any conditions.  The results reported in Figure \ref{fig:ab1} indicate that conditioning F0 on the decoder can improve the speech quality of the converted speech samples.

\begin{figure}[t]
    \centering
    \includegraphics[width=6cm]{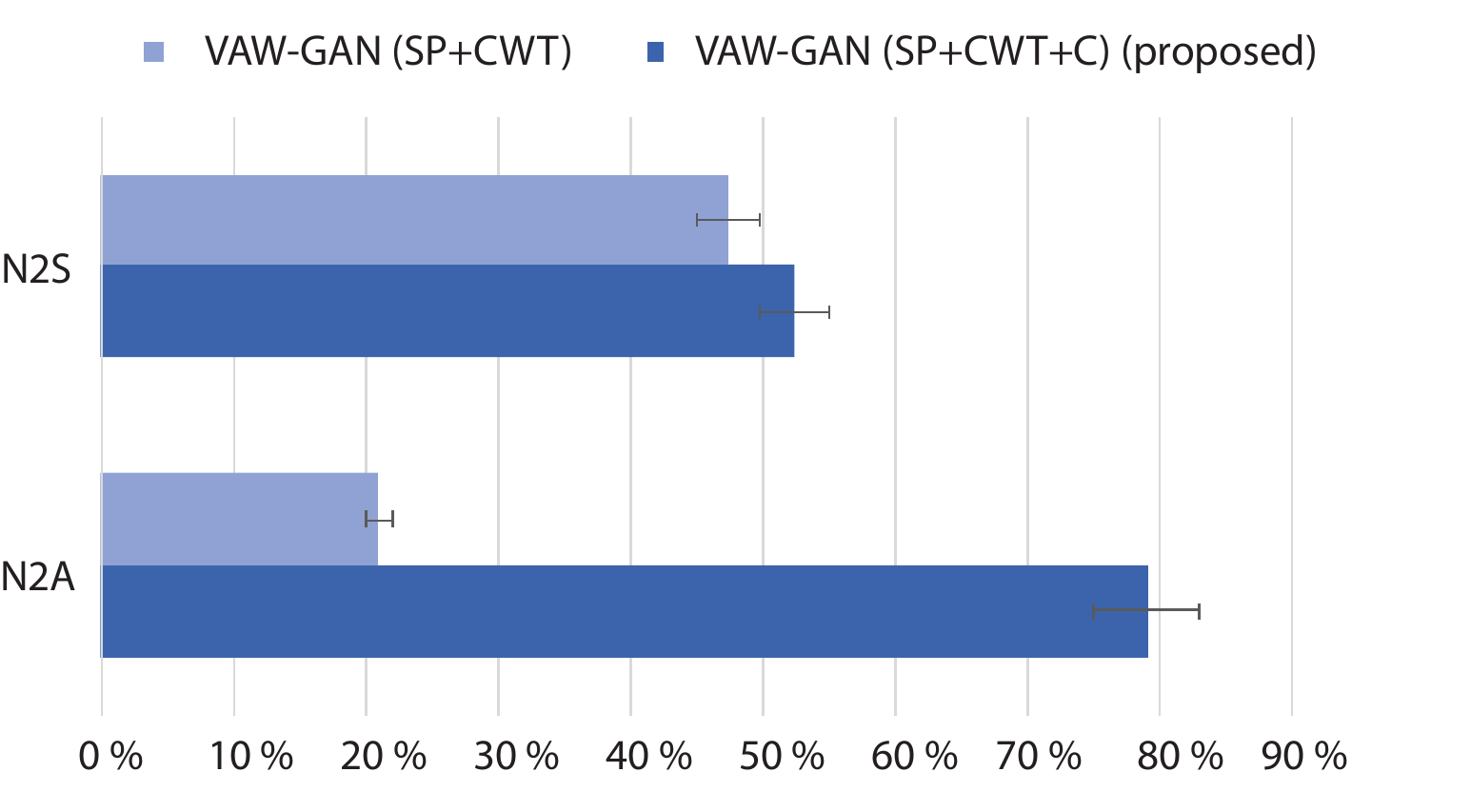}
   \vspace{-4mm}
    \caption{XAB emotion similarity preference results with 95 $\%$ confidence interval to assess the effect of F0 conditioning.}
    \vspace{-4mm}
    \label{fig:ab2}
\end{figure}
\begin{figure}[t]
    \centering
    \includegraphics[width=6cm]{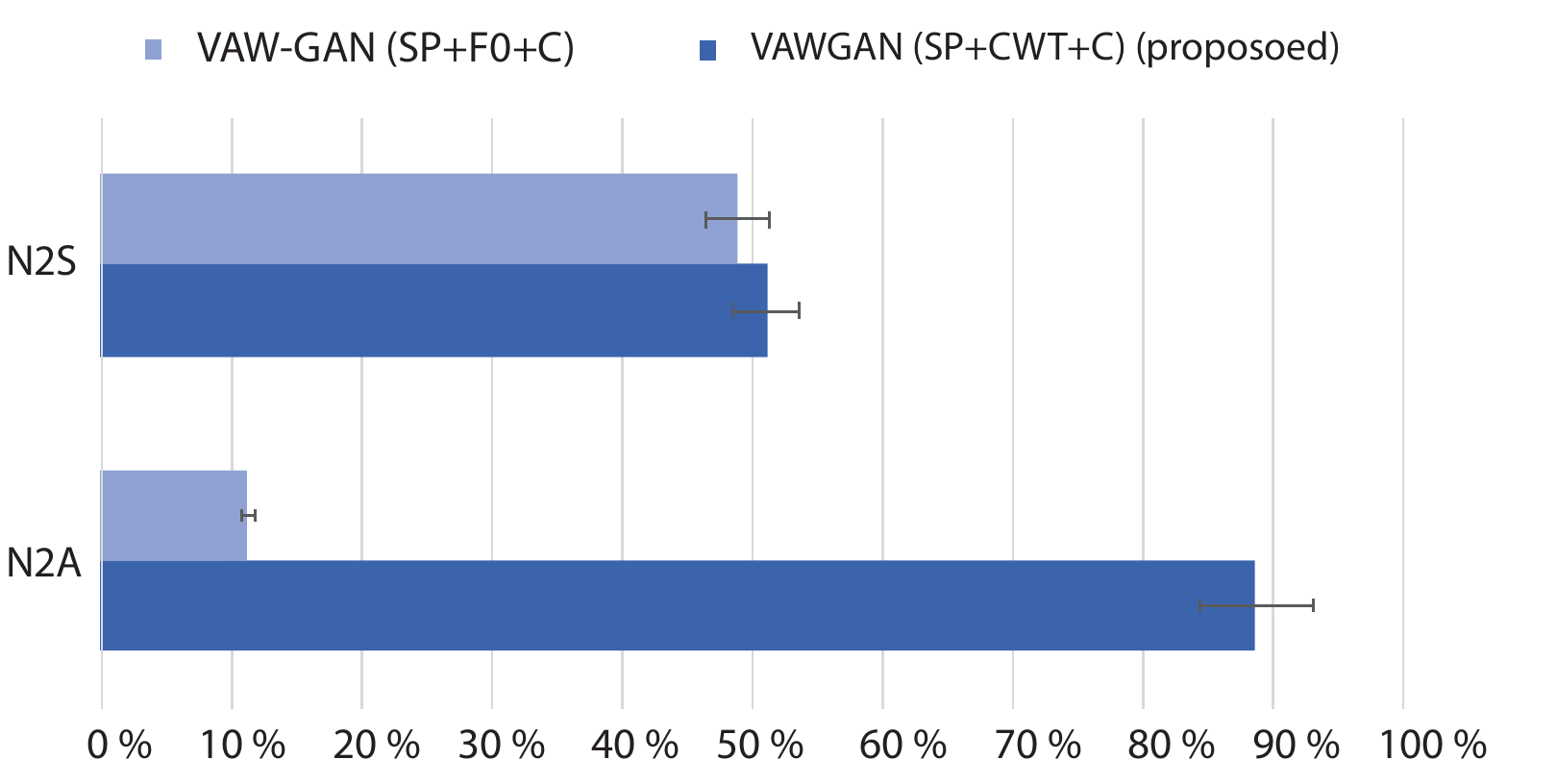}
    \vspace{-4mm}
    \caption{XAB emotion similarity preference results with 95 $\%$ confidence interval to assess the effect of CWT modelling of F0.}
    \label{fig:ab3}
    \vspace{-4mm}
\end{figure}
We then conduct XAB emotion similarity test to assess the performance of emotional expression in the speech. In this experiment, we choose \textit{VAW-GAN (SP+CWT)} as the baseline framework, which is consistent with the speech quality test. All the subjects are asked to choose the one which sounds closer to the reference samples in terms of emotional expression. As reported in Figure \ref{fig:ab2}, we observe that our proposed framework consistently has better performance than the baseline framework for both N2S and N2A, which we believe is remarkable. These results show that frameworks with F0 conditioning can improve the emotion performance compared with those frameworks without any conditions.

We further conduct XAB emotion similarity test to compare the proposed framework with the baseline framework \textit{VAW-GAN (SP+CWT)}. We note that both frameworks condition the decoder with F0 features, whereas the proposed framework converts F0 with CWT modelling and the baseline framework transforms F0 in a traditional manner with LG-based linear transformation method. As shown in Figure \ref{fig:ab3}, our proposed framework shows comparable results with the baseline frameworks in N2S and achieves much better results in N2A. These results indicate that our proposed framework with CWT modelling 
of F0 can significantly improve the emotion performance especially for N2A.


\section{Conclusion}
In this paper, we propose a speaker-dependent emotional voice conversion framework. We perform both spectrum and prosody conversion based on VAW-GAN. We perform a non-linear prosody modelling method which uses CWT to decompose F0 into different time-scales. Moreover, we investigate the effect of  F0 conditioning on the decoder to improve the emotion conversion performance. Experimental results show the proposed framework can achieve better performance than the baseline frameworks with non-parallel training data. In the future work, we will explore to modify energy and duration for emotion conversion.

\section{Acknowledgement}

This work is supported by the National Research Foundation, Singapore under its AI Singapore Programme (AISG Award No: AISG-100E-2018-006, AISG-GC-2019-002),  the National Robotics Programme (Programmatic Grant Number: 192 25 00054), Human Robot Collaborative AI for AME Programmatic Grant (Programmatic Grant No. A18A2b0046) and Neuromorphic Computing Programmatic Grant (Programmatic Grant No. A1687b0033) from the Singapore Government’s Research, Innovation and Enterprise 2020 plan in the Advanced Manufacturing and Engineering domain.
This work is also supported by SUTD Start-up Grant Artificial Intelligence for Human Voice  Conversion  (SRG  ISTD  2020  158)  and  SUTD  AI  Grant  titled 'The Understanding and Synthesis of Expressive Speech by AI' (PIE-SGP-AI-2020-02).

\footnotesize
\bibliographystyle{IEEEbib}
\bibliography{mybib}

\end{document}